# Metallic Phase and Temperature Dependence of the $v = 0$ Quantum Hall State in Bilayer Graphene

Jing Li[1†], Hailong Fu[1], Zhenxi Yin[1], Kenji Watanabe[2], Takashi Taniguchi[2], Jun Zhu[1,3*]

[1]Department of Physics, The Pennsylvania State University, University Park, Pennsylvania 16802, USA.

[2]National Institute for Material Science, 1-1 Namiki, Tsukuba 305-0044, Japan.

[3]Center for 2-Dimensional and Layered Materials, The Pennsylvania State University, University Park, Pennsylvania 16802, USA.

*Correspondence to: jzhu@phys.psu.edu (J. Zhu)

[†]Present address: National High Magnetic Field Laboratory, Los Alamos, NM 87544, USA.

**Abstract**

The $v = 0$ quantum Hall state of bilayer graphene is a fertile playground to realize many-body ground states with various broken symmetries. Here we report the experimental observations of a previously unreported metallic phase. The metallic phase resides in the phase space between the previously identified layer polarized state at large transverse electric field and the canted antiferromagnetic state at small transverse electric field. We also report temperature dependence studies of the quantum spin Hall state of $v = 0$. Complex non-monotonic behavior reveals concomitant bulk and edge conductions and excitations. These results provide timely experimental update to understand the rich physics of the $v = 0$ state.

Bilayer graphene (BLG) in a magnetic field $B$ offers an exciting opportunity to examine the emergence of many-body ground states arising from its multiple internal electronic degrees of freedom. The non-interacting and unbiased $v = 0$ possesses eight approximately degenerate Landau levels (LLs) labeled by their spin ($\downarrow$ and $\uparrow$), valley (K and K′) and orbital ($N = 0, 1$) quantum numbers. The interplay between Coulomb interactions in a perpendicular magnetic field $B_\perp$ and layer/ valley polarization driven by an external electric field $D$ gives rise to many-body ground states with various broken symmetries [1-23]. Effective short-ranged interactions, for example, is shown to stabilize a canted spin anti-ferromagnetic (CAF), layer coherent ground state at small $D$, which transitions to a fully layer (FLP), spin unpolarized state at large $D$ [9,10]. Experiments to date support this scenario, where a conductance peak is commonly associated with the putative CAF/ FLP phase boundary [18,20,23-25]. This line boundary further splits into two branches at high perpendicular magnetic field $B_\perp > 12$ T, with the phase region in between thought to be partially polarized in all indices [20,23,26,27].

Even at a moderate $B_\perp$, calculations including more hopping terms and other symmetry-allowed electron-electron interaction terms have uncovered a more nuanced picture [11,17,26]. For example, theory shows that a broken-U (1) × U (1) phase that is canted in both spin and valley could be stabilized in certain parameter regimes in the vicinity of the CAF/ FLP phase boundary [26]. Their experimental plausibility remains to be tested.

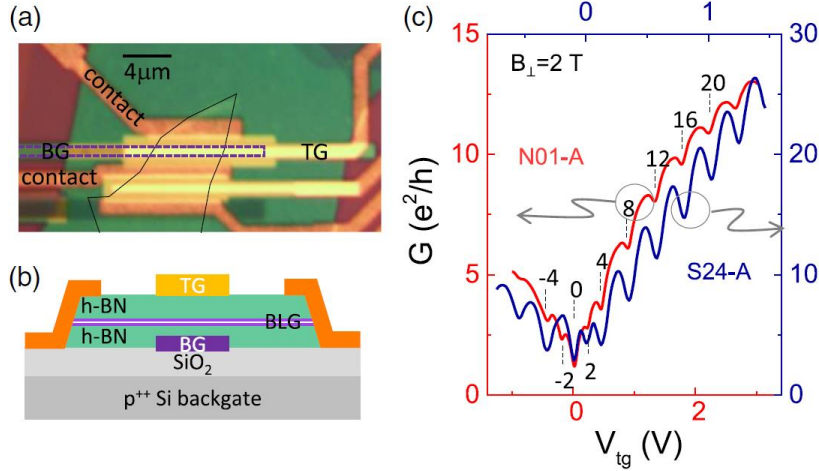

FIG.1. (a) The optical micrograph of device N01-A. The thin black line outlines the BLG sheet. The local bottom gate (BG) is outlined in dashed purple. The top gate (TG) overlaps the BG over the BLG sheet. The two side contacts are as labeled. (b) A schematic side view of the device. The TG and side contacts are made of Cr/ Au and the BG is made of multi-layer graphite. (c) Two terminal conductance $G$ ($V_{tg}$) in devices N01-A (red trace and left and bottom axes) and S24-A (blue trace and top and right axes) at $B_\perp = 2$ T. $V_{bg} = 0$ V, $V_{Si} = 40$ V, and $T = 0.3$ K for both traces. The filling factors are marked in the figure.

The ground state of $v = 0$ becomes yet richer when the spin polarization is explicitly controlled. Previous studies showed that the CAF state undergoes a second-order phase transition to a spin ferromagnet (FM) upon the application of a large in-plane magnetic field $B_{//}$. The FM state is a bulk insulator but possesses metallic quantum spin Hall (QSH) edge states [18,21]. The QSH edge states are a promising platform to examine the interactions of helical Luttinger liquid [28], and explore the potential realization of an interaction-driven bosonic symmetry protected topological state [29,30]. The pursuit of these intriguing possibilities requires substantial experimental knowledge of the FM (QSH) state, which is currently lacking.

The purpose of this Letter is two-fold. We first report the observation of a previously unreported metallic (M) phase of $v = 0$. The M phase is marked by very sharp conductance changes and occupies a large phase space in the vicinity of the previously observed CAF/ FLP phase boundary. We discuss its possible origins. The M phase persists in a large $B_{//}$, with a temperature dependence that is distinct from the CAF and FM phases. We also conducted a systematic study of the temperature-dependent conductance of the FM phase. The results indicate contributions from both bulk and edge conducting channels and their temperature-dependent excitations. Our results provide fresh input to further understand the fascinating behavior of the $v = 0$ state in BLG.

Our measurements employ two-terminal, dual-gated, hexagonal boron nitride (h-BN) encapsulated BLG devices, an exemplary optical micrograph and schematic side view of which are shown in Figs. 1(a) and 1(b). The fabrication of the devices uses van der Waals transfer stacking, side contacts, and a multi-layer graphite local bottom gate to screen charged impurities [23,31]. The dual-gated region is connected to the side contacts through regions heavily doped by the Si backgate. The total resistance outside the dual-gated region is less than 300 Ω (See Fig. S2) and not subtracted from measurements. Our devices are in a "wide bar" geometry as Fig. 1(a) shows. They are separated into two groups labeled as "standard" (S) or "new" (N). Figure 1(c) compares the two-terminal conductance of devices N01-A and S01-A at magnetic field $B = 2$ T. Both exhibit strong quantum Hall effect including the appearance of $v = \pm 2$ at this low $B$-field, indicating high sample quality. (See Section 1 of the Supplementary Information for the fabrication and characteristics of the devices.)

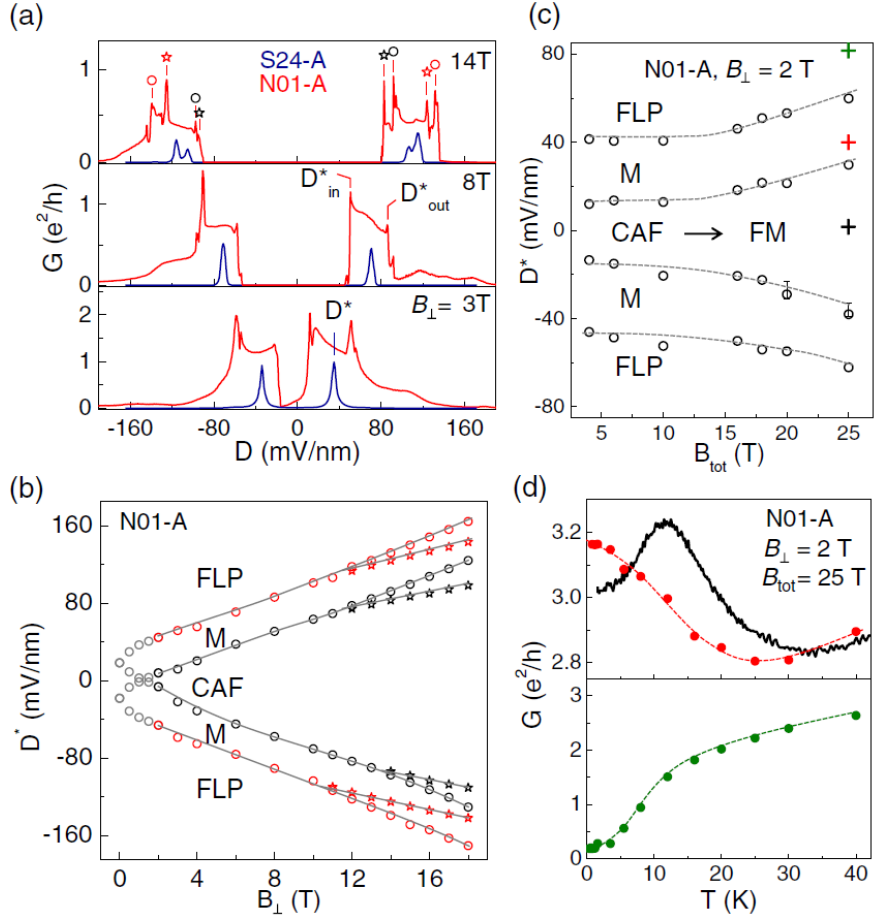

FIG.2. (a) Comparison of $G(D)$ in device N01-A (red traces) and S24-A (blue traces) at $B_\perp$ =3, 8, 14 T. Thin vertical lines and symbols mark $D^*$ in S24-A and $D^*_{in/out}$ in N01-A. $D^*_{in/out}$ follow a prominent peak consistently as $B_\perp$ evolves. $T = 20$ mK for the red traces and $T = 1.4$ K for the blue traces. (b) ($D^*$, $B_\perp$) phase diagram of N01-A constructed from data in (a) and additional data given in Figs. S3 and S4. Below 2 T, boundary conductance peaks (open gray circles) are present although the conductance of the CAF phase does not drop to zero (See Fig. S3). Gray lines are guide to the eye (c) The ($D^*$, $B_{tot}$) phase diagram of N01-A constructed from data in Fig. S7. Dashed lines are guide to the eye. The three "+" symbols mark the spots where $T$-dependent conductance measurements were taken. (d) plots the $G(T)$ data. $D^* = 0$, 38.2, and 80.1 mV/nm respectively for the black, green and red spots. $B_\perp = 2$ T and $B_{tot} = 25$ T for all three.

We obtain the displacement field $D$ dependence of the two-terminal conductance $G(D)$ by sweeping the top and bottom gates (TG and BG) simultaneously while maintaining $v = 0$ [32,33] (See Section 1 of the Supplementary Information for a description of methods). Figure 2(a) compares $G(D)$ obtained in devices S24-A and N01-A at selected $B_\perp$ fields. S24-A (blue traces) exhibits the "standard" behavior, where the CAF/ FLP phase transition is marked by a single conductance peak labeled as $D^*$ in the plot. In device N01-A (red traces), this peak is widened to a region of high conductance defined by two peaks labeled as $D^*_{in}$ and $D^*_{out}$. Both $D^*$ and $D^*_{in/out}$ increase with increasing $B$, each peak splitting into two at $B_\perp > 11$ T. In the top panel of Fig. 2(a), we mark the four conductance peaks in N01-A using stars and circles. The conductance of regions in between slowly decreases with increasing $B_\perp$ but remains substantial (See Figs. S3 - S4 for additional $G(D, B)$ data in N01-A in the field range of 0 - 18 T). The lack of a clear temperature dependence at temperatures below 1 K is consistent with a metallic nature of the state. (See Fig. S5 for data). By tracking the $B_\perp$ evolution of the conductance peaks, we have constructed a new phase diagram for N01-A and plotted it in Fig. 2(b). This phase diagram is reproducible from cool-down to cool-down and is also observed in a second device N01-B (See Fig. S6). The most remarkable feature of this phase diagram is the prominent appearance of the metallic region. The M phase persists when an in-plane magnetic field component $B_{//}$ is added. Figure S7 plots $G(D)$ at a fixed $B_\perp = 2$ T but with increasing total field $B_{tot}$ up to 25 T. Here $B_{tot}^2 = B_{//}^2 + B_\perp^2$. The conductance of the M phase increases with increasing $B_{tot}$ and reaches about 3.2 $e^2/h$ at $B_{tot} = 25$ T. Figure 2(c) plots the phase diagram in the ($D^*$, $B_{tot}$) plane constructed from data in Fig. S7. The width of the M phases in $D$ appears to be approximately constant on this diagram. Figure 2(d) compares the $T$-dependent conductance $G(T)$ at three locations marked by the three "+" symbols. The state inside the M phase (red trace) exhibits a clear metallic behavior below 25 K. The state inside the FLP phase (green trace) is insulating while the state inside the FM phase (black trace) exhibits a non-monotonic $G(T)$, which we will discuss later. The behavior of the M phase in a large $B_{tot}$ imposes additional constraints on its potential interpretations.

To understand the origin of the M phase, we examined several scenarios. First of all, device N01-A exhibits a small carrier density broadening of $\delta n \sim 6 \times 10^9$ cm$^{-2}$ (Fig. S2) and very abrupt transitions of $\delta D \sim 2$ mV/nm in Fig. 2(a). The high sample qualities rule out disorder broadening as the source of the wide high-conductance region. Comparing the phase diagrams of the S-type and N-type devices, we see that *superficially* the "new" phase diagram resembles the "sum" of two "standard" phases diagrams, offset by $\Delta D \sim 40$ mV/nm from one another. One hypothetic scenario would be that somehow the device consisted of two areas experiencing different $D$-fields and the M region is composed of CAF and FLP states existing in different areas of the sample. This is a highly unusual experimental situation, since a net change in $D$, *without change in n*, requires concerted unintentional doping from the top and bottom interfaces. Examining the fabrication of both S- and N-type devices carefully we could not see how such situation could occur in our devices. Nor has it been reported by other groups. Nonetheless we have simulated in Fig. S8 the total conductance of a device consisting of two areas experiencing different $D$-fields and show that it could not explain the high conductance of the M phase.

In the "standard" devices, the conductance peak at the first-order CAF/ FLP transition ($D^*$) in Fig. 2(a) is attributed to a microscopic network of CAF and FLP phases [18,20,23-25]. Applied to the M phase, this network model would imply that the CAF/ FLP phase transition occur over an extended range of $D$-field, which seems highly unusual for first-order transitions. In the literature, an insulator-metallic-insulator phase transition was proposed to explain the experimentally observed nonscaling behavior in the quantum Hall regime [34]. However, that is an unusual case with physics specific to the quantum Hall systems.

Alternatively, the M phase could be a new ground state with a different broken symmetry. Candidates include, for example, a spin-valley entangled phase and a broken-U (1) × U (1) phase as discussed by Murthy et al. [26]. The splitting at higher $B_\perp$ can also in principle arise from additional predicted phases [26]. These new broken symmetry states produce conductance broadly consistent with our observations and we hope that our results stimulate more studies to directly probe their order parameters. Towards this goal, we note that the h-BN encapsulation layers used in our N devices (56/ 57 nm, See Table S1) is much thicker than a typical thickness of 15 - 30 nm used by us and others. Whether and how this impacts the ground state of $v = 0$ require more systematic studies to clarify. It is also worth noting that suspended BLG devices also exhibit both the "standard" (Fig. 2 of Ref. [5]) and "new" (Fig. S6 of Ref. [12]) types of CAF/FLP transitions. Additionally, a hint of the metallic phase was perhaps present in Fig. 2(c) of Ref. [18] though stronger disorder made the situation less clear.

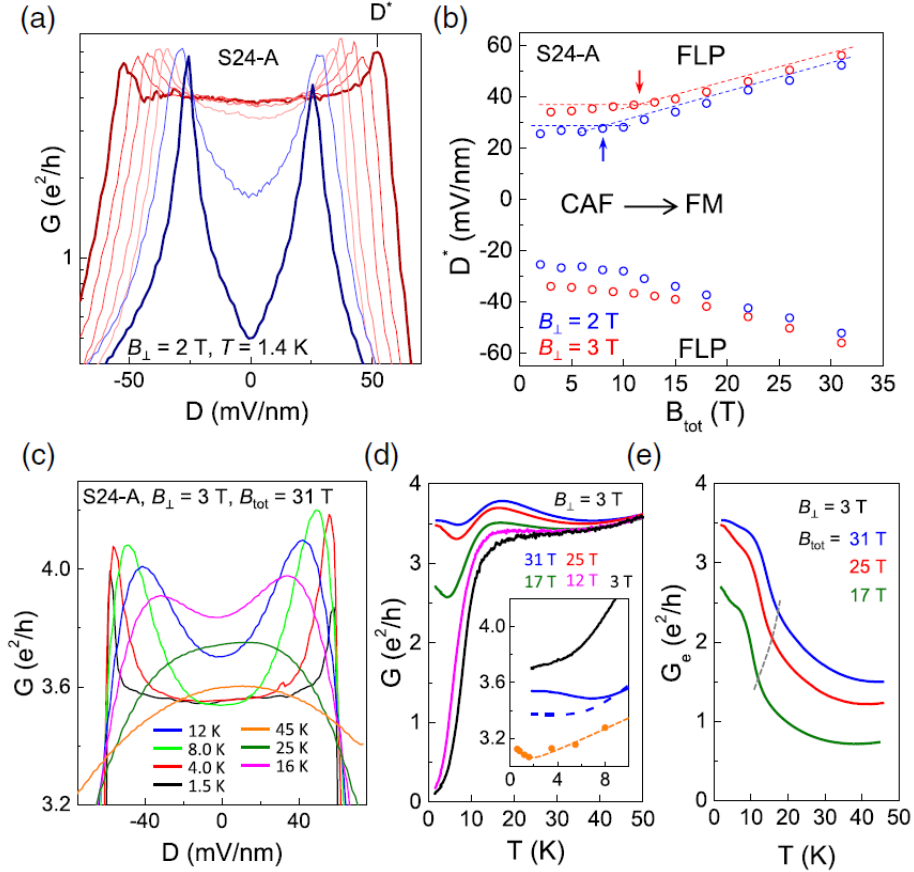

FIG. 3. (a) $G(D)$ of $v = 0$ in device S24-A at $B_\perp = 2$ T and selected $B_{tot}$'s. From blue to red: $B_{tot} = 2, 10, 15, 18, 22, 26$ and 31T. (b) $(D^*, B_{tot})$ phase diagram constructed by reading transition $D^*$, as marked in (a). $B_\perp = 2$ (3) T for blue (red) symbols. The arrows point to intersections of low-$D$ and high-$D$ trend lines, which are $(B^*_{tot}, D^*) = (8.3$ T, 26 mV/nm) and (11.2 T, 34 mV/nm) respectively for $B_\perp = 2$ T and 3 T. (c) $G(D)$ of the same device at selected temperatures as indicated in the plot. $B_\perp = 3$ T and $B_{tot} = 31$ T. (d) $G(T)$ at $D = 0$. From top to bottom $B_{tot} = 31, 25, 17, 12, 3$ T. $B_\perp = 3$ T for all traces. Inset: data in the low-$T$ range. The blue solid and dashed traces correspond to $(B_{tot}, B_\perp) = (31, 3)$ T in two different cool-downs. The black trace corresponds to $(B_{tot}, B_\perp) = (31, 2)$ T. They are all taken on S24-A. The orange data is from device N01-A at $(B_{tot}, B_\perp) = (25, 2)$ T. The sign of $dG/dT$ at low temperatures varies. (e) Remaining $G_e(T)$ of top three traces in (d) after subtracting "bulk contribution" as described in the text. The subtracted data are given in Fig. S10. The dashed line marks the estimated temperatures, below which $G_e(T)$ rises rapidly.

We now turn to the temperature dependence of the $v = 0$ state, where experimental studies were spotty [18,21]. Figure 3(a) plots a set of $G(D)$ traces in device S24-A at $B_\perp = 2$ T and selected $B_{tot}$'s ranging 2 - 31 T. The saturation of $G(0) = 3.9$ $e^2/h$ at $B_{tot} \geq 18$ T supports the $B_{//}$-driven crossover to a bulk FM with QSH edge states, as reported in Maher et al [18]. Here, conductance peaks marking the FM/ FLP transitions ($D^*$ in Fig. 3(a)) are sharp and enable us to determine the phase boundary accurately. Figure 3(b) plots two phase diagrams we obtained at $B_\perp = 2$ T and 3 T respectively, as a function of $B_{tot}$. Following Ref. [18], we identify the crossover field $B^*_{tot}$ as the point where the low and high-$D$ trend lines intersect. $B^*_{tot}$ is respectively 8.3 T/ 11.2 T for $B_\perp = 2$ T/ 3 T, which are considerably smaller than values reported in Ref. [18]. We associate the FM/FLP phase boundary with the condition, where the increment of the Zeeman energy $E_z = g\mu_B B_{tot}$ equates the increment of the valley splitting energy $\Delta_v$. Using $\Delta_v = 0.13$ $D$ determined previously in Ref. [23] and the slope of $dD^*/dB_{tot} = 1.0$ mV/nm/T in Fig. 3(b), we obtain a $g$-factor of 2.2. This value is in excellent agreement with the expectation of nearly free electrons in graphene.

Next, we investigate the temperature dependence of the CAF and FM states. Figure 3(c) plots an example of $G(D)$ in the FM phase at selected temperatures. As $T$ increases, the sharp conductance peaks at $\pm D^*$ become increasingly broadened and indistinguishable at $T^* \sim 25$ K. Meanwhile, the small $D$ region occupied by the FM phase remains metallic and its transition to the insulating FLP phase at large $D$ is well-defined by a "bundle" point. This suggests that the FM phase melts to a metallic state at high temperature [35]. Indeed, in a single-particle LL picture, the $\nu = 0$ state is a half-filled, metallic QH liquid [1,35]. $T^*$ thus provides a formation energy scale of the FM phase, which was not determined previously. It is a fraction of the single-particle gap to the $N = 2$ LL, which is about 170 K using band parameter determined in Ref. [36]. This melting transition is difficult to identify in conductance change, since both the edge states and the half-filling QH liquid exhibit similar conductance of $\sim 4$ $e^2/h$.

$G(T)$ at varying $B_\perp$ and $B_{tot}$ provides a closer look. Figure 3(d) plots the $D = 0$ $G(T)$ in S24-A at $B_\perp = 3$ T and selected $B_{tot}$'s ranging 3 - 31 T. In the high temperature limit, all five traces merge and exhibit a slightly positive $dG/dT$, which is consistent with the temperature dependence of a half-filled QH liquid. As $T$ decreases and the CAF or FM state forms, $G(T)$ starts to exhibit diverse behaviors. The two lower traces correspond to the CAF state, where the bulk conduction continues to decrease with decreasing temperature (See Fig. S7 for more data and discussions of the CAF state). In the top three traces ($B_{tot} = 17 - 31$ T), the QSH edge states are developed and the measured $G(T)$ is high but also non-monotonic. We have verified that this $T$-dependence originates from the dual-gated area, i.e. intrinsic to the $\nu = 0$ state, and not from the contacts or the access region (See discussions of Fig. S2). A potential explanation arises, by assuming that the bulk of the FM phase exhibits behavior similar to that of the CAF phase, and that the measured $G(T)$ is a sum of the bulk and edge contributions. In this picture, it is perhaps not difficult to see that $dG/dT$ can change sign in different temperature ranges, in sample-dependent and condition-dependent manners. The several data sets we obtained in both "new" and "standard" devices and at different $B_\perp$'s are all consistent with this bulk + edge parallel conduction scenario. Specifically the temperature dependence of the conductance is quite diverse at the lowest temperatures, as a few examples plotted in the inset of Fig. 3(d) show. The dimensions of the device also play a role in the quantitative behavior of $G(T)$. (See Fig. S9 for more discussions)

Finally, we performed a crude "background subtraction" process to estimate the conduction of the edge states in the FM (QSH) phase. For each targeted FM state, we first identify an effective field $B_\perp^{\text{eff}}$, at which the FLP/ CAF transition occurs at the same $D^*$ as the FLP/ FM transition of the targeted FM (QSH) state, using the ($D^*$, $B_\perp$) and ($D^*$, $B_{\text{tot}}$) phase diagrams of the device together. We obtained $B_\perp^{\text{eff}}$ = 5.9, 4.9, and 3.7 T respectively for the top three curves in Fig. 3(d), where $B_\perp$ = 3 T and $B_{\text{tot}}$ = 31 T, 25 T and 17 T respectively. We then approximate the bulk $G$ ($T$) of the FM state by that of the corresponding CAF state--measurement and discussions are given in Fig. S10 of the supplementary material--and substrate it from the measured total conductance in Fig. 3(d). The remaining $G_e$ ($T$) is attributed to the edge states. Figure 3(e) plots $G_e$ ($T$) of the top three curves in the main panel of Fig. 3(d). The conductance of the edge states rises rapidly with decreasing temperature below 15 - 20 K. This behavior has a natural explanation. Because the backscattering of the QSH edge states involves spin flips, one way this process can be mediated is through spin wave excitations of the bulk [37]. As temperature decreases, such excitations become increasingly suppressed, hence the edge state conductance increases. In addition, the negative d$G$/ d$T$ is also consistent with the temperature dependence predicted for a repulsive helical Luttinger liquid in certain temperature range [37-39]. The differentiation of these two mechanisms will require more studies. Although both scenarios expect a negative d$G$/ d$T$ at low temperature, our observations are more diverse, as the inset of Fig. 3(d) shows. It is possible that varying bulk conductions contribute to such non-universal behavior but we also cannot rule out intrinsic behavior of the edge states manifesting differently under different interaction conditions, e.g. different $B_\perp$'s. The complex conductance behavior our measurements revealed makes it clear that understanding the conduction of the bulk FM state is a necessary part of probing the intrinsic behavior of the QSH edge states. Measurements at lower temperature using non-local transport geometries [40,41] or controlled tunneling at a quantum point contact will be essential to further explore the exciting prospects of the edge states in bilayer graphene [28-30].

In summary, our systematic study of the $v$ = 0 quantum Hall state in BLG has produced new information and insights on this fertile many-body platform. A metallic phase is found in the vicinity of the previously reported CAF/ FLP phase boundary, with potential implications for new broken symmetry ground states. The FM (QSH) state is shown to exhibit complex temperature dependence, likely due to significant concomitant contributions from both bulk and edge conductions. We hope that our results stimulate more future theoretical and experimental studies.

**Acknowledgement**
Work at Penn State is supported by the NSF through NSF-DMR-1506212 and NSF-DMR-1708972. Work atNIMS is supported by the Elemental Strategy Initiative conducted by the MEXT, Japan and the CREST (JPMJCR15F3), JST. Part of this work was performed at the NHMFL, which was supported by the NSF through NSF-DMR-1157490, NSF-DMR-0084173 and the State of Florida. We thank Herbert Fertig, Efrat Shimshoni, Ganpathy Murthy, Jainendra Jain, Chaoxing Liu, Chun Ning Lau and Philip Kim for helpful discussions and Jan Jaroszynski of the NHMFL for experimental assistance.

**Reference**

[1]     E. McCann and V. Fal'ko, Landau-level degeneracy and quantum hall effect in a graphite bilayer. *Physical Review Letters* **96**, 086805 (2006).

[2] E. V. Castro, N. M. R. Peres, T. Stauber, and N. A. P. Silva, Low-Density Ferromagnetism in Biased Bilayer Graphene. *Physical Review Letters* **100**, 186803 (2008).
[3] B. E. Feldman, J. Martin, and A. Yacoby, Broken-symmetry states and divergent resistance in suspended bilayer graphene. *Nature Physics* **5**, 889 (2009).
[4] R. Nandkishore and L. Levitov, Quantum anomalous Hall state in bilayer graphene. *Physical Review B* **82**, 115124 (2010).
[5] R. T. Weitz, M. T. Allen, B. E. Feldman, J. Martin, and A. Yacoby, Broken-Symmetry States in Doubly Gated Suspended Bilayer Graphene. *Science* **330**, 812 (2010).
[6] Y. Zhao, P. Cadden-Zimansky, Z. Jiang, and P. Kim, Symmetry Breaking in the Zero-Energy Landau Level in Bilayer Graphene. *Physical Review Letters* **104**, 066801 (2010).
[7] O. Vafek and K. Yang, Many-body instability of Coulomb interacting bilayer graphene: Renormalization group approach. *Physical Review B* **81**, 041401 (2010).
[8] Y. Lemonik, I. L. Aleiner, C. Toke, and V. I. Fal'ko, Spontaneous symmetry breaking and Lifshitz transition in bilayer graphene. *Physical Review B* **82**, 201408 (2010).
[9] M. Kharitonov, Antiferromagnetic state in bilayer graphene. *Physical Review B* **86**, 195435 (2012).
[10] M. Kharitonov, Canted Antiferromagnetic Phase of the $\nu = 0$ Quantum Hall State in Bilayer Graphene. *Physical Review Letters* **109**, 046803 (2012).
[11] Y. Lemonik, I. Aleiner, and V. I. Fal'ko, Competing nematic, antiferromagnetic, and spin-flux orders in the ground state of bilayer graphene. *Physical Review B* **85**, 245451 (2012).
[12] J. Velasco Jr, L. Jing, W. Bao, Y. Lee, P. Kratz, V. Aji, M. Bockrath, C. N. Lau, C. Varma, R. Stillwell, D. Smirnov, F. Zhang, J. Jung, and A. H. MacDonald, Transport spectroscopy of symmetry-broken insulating states in bilayer graphene. *Nature Nanotechnology* **7**, 156 (2012).
[13] A. Veligura, H. J. van Elferen, N. Tombros, J. C. Maan, U. Zeitler, and B. J. van Wees, Transport gap in suspended bilayer graphene at zero magnetic field. *Physical Review B* **85**, 155412 (2012).
[14] W. Bao, J. Velasco, F. Zhang, L. Jing, B. Standley, D. Smirnov, M. Bockrath, A. H. MacDonald, and C. N. Lau, Evidence for a spontaneous gapped state in ultraclean bilayer graphene. *Proceedings of the National Academy of Sciences* **109**, 10802 (2012).
[15] F. Zhang and A. H. MacDonald, Distinguishing Spontaneous Quantum Hall States in Bilayer Graphene. *Physical Review Letters* **108**, 186804 (2012).
[16] F. Freitag, J. Trbovic, M. Weiss, and C. Schönenberger, Spontaneously Gapped Ground State in Suspended Bilayer Graphene. *Physical Review Letters* **108**, 076602 (2012).
[17] J. Lambert and R. Côté, Quantum Hall ferromagnetic phases in the Landau level $N = 0$ of a graphene bilayer. *Physical Review B* **87**, 115415 (2013).
[18] P. Maher, C. R. Dean, A. F. Young, T. Taniguchi, K. Watanabe, K. L. Shepard, J. Hone, and P. Kim, Evidence for a spin phase transition at charge neutrality in bilayer graphene. *Nature Physics* **9**, 154 (2013).
[19] A. Young, J. Sanchez-Yamagishi, B. Hunt, S. Choi, K. Watanabe, T. Taniguchi, R. Ashoori, and P. Jarillo-Herrero, Tunable symmetry breaking and helical edge transport in a graphene quantum spin Hall state. *Nature* **505**, 528 (2013).
[20] K. Lee, B. Fallahazad, J. Xue, D. C. Dillen, K. Kim, T. Taniguchi, K. Watanabe, and E. Tutuc, Chemical potential and quantum Hall ferromagnetism in bilayer graphene. *Science* **345**, 58 (2014).

[21] S. Pezzini, C. Cobaleda, B. A. Piot, V. Bellani, and E. Diez, Critical point for the canted antiferromagnetic to ferromagnetic phase transition at charge neutrality in bilayer graphene. *Physical Review B* **90**, 121404 (2014).
[22] Y. H. Wu, T. Shi, and J. K. Jain, Non-Abelian Parton Fractional Quantum Hall Effect in Multilayer Graphene. *Nano Letters* **17**, 4643 (2017).
[23] J. Li, Y. Tupikov, K. Watanabe, T. Taniguchi, and J. Zhu, Effective Landau Level Diagram of Bilayer Graphene. *Physical Review Letters* **120**, 047701 (2018).
[24] T. Jungwirth and A. H. Macdonald, Resistance spikes and domain wall loops in Ising quantum Hall ferromagnets. *Physical Review Letters* **87**, 216801 (2001).
[25] K. Dhochak, E. Shimshoni, and E. Berg, Spontaneous layer polarization and conducting domain walls in the quantum Hall regime of bilayer graphene. *Physical Review B* **91**, 165107 (2015).
[26] G. Murthy, E. Shimshoni, and H. A. Fertig, Spin-valley coherent phases of the $v = 0$ quantum Hall state in bilayer graphene. *Physical Review B* **96**, 245125 (2017).
[27] B. M. Hunt, J. I. A. Li, A. A. Zibrov, L. Wang, T. Taniguchi, K. Watanabe, J. Hone, C. R. Dean, M. Zaletel, R. C. Ashoori, and A. F. Young, Direct measurement of discrete valley and orbital quantum numbers in bilayer graphene. *Nature Communications* **8**, 948 (2017).
[28] J. C. Y. Teo and C. L. Kane, Crtical behavior of a point contact in a quantum spin Hall insulator. *Physical Review B* **79**, 235321 (2009).
[29] Z. Bi, R. Zhang, Y.-Z. You, A. Young, L. Balents, C.-X. Liu, and C. Xu, Bilayer Graphene as a Platform for Bosonic Symmetry-Protected Topological States. *Physical Review Letters* **118**, 126801 (2017).
[30] R.-X. Zhang and C.-X. Liu, Fingerprints of a Bosonic Symmetry-Protected Topological State in a Quantum Point Contact. *Physical Review Letters* **118**, 216803 (2017).
[31] J. Li, R. X. Zhang, Z. Yin, J. Zhang, K. Watanabe, T. Taniguchi, C. X. Liu, and J. Zhu, A valley valve and electron beam splitter. *Science* **362**, 1149 (2018).
[32] J. Li, K. Wang, K. J. McFaul, Z. Zern, Y. Ren, K. Watanabe, T. Taniguchi, Z. Qiao, and J. Zhu, Gate-controlled topological conducting channels in bilayer graphene. *Nature Nanotechnology* **11**, 1060 (2016).
[33] J. Li, H. Wen, K. Watanabe, T. Taniguchi, and J. Zhu, Gate-Controlled Transmission of Quantum Hall Edge States in Bilayer Graphene. *Physical Review Letters* **120**, 057701 (2018).
[34] G. Xiong, S.-D. Wang, Q. Niu, D.-C. Tian, and X. R. Wang, Metallic Phase in Quantum Hall Systems due to Inter-Landau-Band Mixing. *Physical Review Letters* **87**, 216802 (2001).
[35] The CAF state in a purely perpendicular magnetic field also melts at comparble temperatures. See Fig. S10 and discussions.
[36] K. Zou, X. Hong, and J. Zhu, Effective mass of electrons and holes in bilayer graphene: Electron-hole asymmetry and electron-electron interaction. *Physical Review B* **84**, 085408 (2011).
[37] Tikhonov, E. Shimshoni, H. A. Fertig, and G. Murthy, Emergence of helical edge construction in graphene at the $v = 0$ quantum hall state. *Physical Review B* **93**, 115137 (2016).
[38] M. Kharitonov, S. Juergens, and B. Trauzettel, Interplay of topology and interactions in quantum Hall topological insulators: U(1) symmetry, tunable Luttinger liquid, and interaction-induced phase transitions. *Physical Review B* **94**, 035146 (2016).
[39] H. A. Fertig and L. Brey, Luttinger Liquid at the Edge of Undoped Graphene in a Strong Magnetic Field. *Physical Review Letters* **97**, 116805 (2006).
[40] A. Roth, C. Brune, H. Buhmann, L. W. Molenkamp, J. Maciejko, X. L. Qi, and S. C. Zhang, Nonlocal Transport in the Quantum Spin Hall State. *Science* **325**, 294 (2009).

[41]    L. Du, I. Knez, G. Sullivan, and R.-R. Du, Robust Helical Edge Transport in Gated InAs /GaSb Bilayers. *Physical Review Letters* **114**, 096802 (2015).

[42]    L. Wang, I. Meric, P. Y. Huang, Q. Gao, Y. Gao, H. Tran, T. Taniguchi, K. Watanabe, L. M. Campos, D. A. Muller, J. Guo, P. Kim, J. Hone, K. L. Shepard, and C. R. Dean, One-Dimensional Electrical Contact to a Two-Dimensional Material. *Science* **342**, 614 (2013).

[43]    Y. Zhang, T.-T. Tang, C. Girit, Z. Hao, M. C. Martin, A. Zettl, M. F. Crommie, Y. R. Shen, and F. Wang, Direct observation of a widely tunable bandgap in bilayer graphene. *Nature* **459**, 820 (2009).

# Supplemental Material for

**Metallic Phase and Temperature Dependence of the *v* = 0 Quantum Hall State in Bilayer Graphene**

Jing Li[1†], Hailong Fu[1], Zhenxi Yin[1], Kenji Watanabe[2], Takashi Taniguchi[2], Jun Zhu[1,3*]

[1]Department of Physics, The Pennsylvania State University, University Park, Pennsylvania 16802, USA.

[2]National Institute for Material Science, 1-1 Namiki, Tsukuba 305-0044, Japan.

[3]Center for 2-Dimensional and Layered Materials, The Pennsylvania State University, University Park, Pennsylvania 16802, USA.

*Corresponding author: jzhu@phys.psu.edu

[†]Present address: National High Magnetic Field Laboratory, Los Alamos, NM 87544, USA.

**Online Supplementary Material Content**

1. Fabrication, measurement and characteristics (Figs S1, S2 and Table S1)
2. Additional data on the metallic phase (Figs. S3 - S8)
3. Additional data on the temperature dependence of the FM and CAF phases (Figs. S9 - S10)

# 1. Fabrication, measurement and characteristics

We begin the fabrication process by exfoliating a graphite sheet of 3 – 5 nm in thickness to $SiO_2$/heavily hole doped silicon substrate. It is then patterned, through e-beam lithography and reaction ion etching into a 1 - 3 μm wide, and 10 - 20 μm long stripe, which serves as the local bottom gate. An h-BN/ BLG/ h-BN stack is assembled using the van der Waals transfer method [42] and transferred onto the graphite bottom gate. The graphite bottom gate is only partially covered by the stack. Cr/ Au side contacts are made to contact the BLG sheet while planar contacts are made to the graphite bottom gate. A top gate (Cr/ Au) is patterned to match the width and position of the local bottom gate. An optical micrograph of device N01-A is shown in Fig. 1(a) of the text. The dimensions of all four devices used in this study are given in Table S1. Figure S2 shows the characteristics of device N01-A.

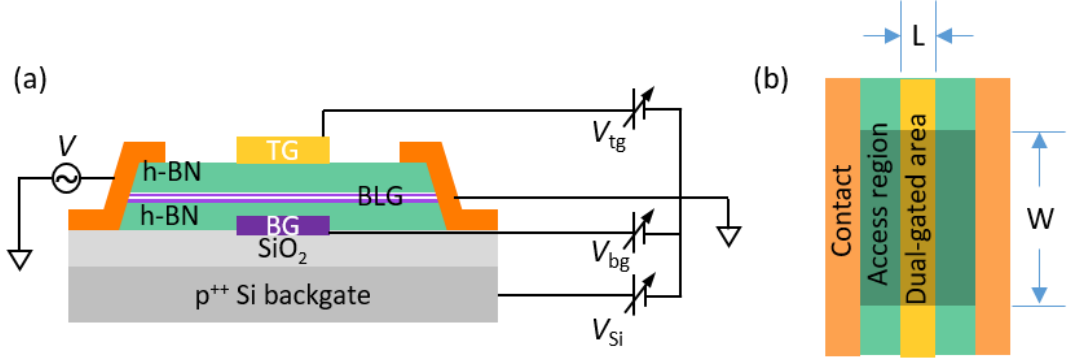

FIG. S1. Schematics of device and measurement. (a) Measurement setup. We carry out standard low-frequency (47 Hz) lock-in measurements. When the resistance is not too high (< ~ 1 MΩ), we pass a constant current *I* through our two-terminal device and measure the voltage *V* to obtain the two-terminal conductance *G* = *I* / *V*. For large resistances (> ~ 1 MΩ), we apply a constant voltage and measure current. We use sufficiently small current or voltage excitations so that the effect of Joule heating is negligible while maintaining a sufficient signal to noise ratio. Silicon gate voltage is fixed at 40 V to heavily dope the region outside the dual-gated area, labeled as access region in (b). We sweep the top and bottom gate voltages simultaneously to change the external displacement field (*D*) while keeping the carrier density at *n* = 0. Following the convention of the field [43], we define $D = \frac{1}{2}\left(\frac{\varepsilon}{d_b}(V_{bg} - V_{bg0}) - \frac{\varepsilon}{d_t}(V_{tg} - V_{tg0})\right)$ where *ε* = 3 is the dielectric constant of h-BN and $d_t$ ($d_b$) is the thickness of the top (bottom) h-BN sheet. $V_{tg0}$ ($V_{bg0}$) is the offset in the top (bottom) gate voltage due to unintentional doping. For more details of the dual gate operation, see Ref. [32, 33].

Table S1: Device Dimensions

| device # | L (um) | w (um) | top hBN thickness | bottom hBN thickness | Presence of M phase |
|---|---|---|---|---|---|
| N01-A | 1 | 6 | 56nm | 57nm | YES |
| N01-B | 3 | 6.6 | 56nm | 57nm | |
| S24-A | 1 | 7 | 28nm | 26nm | NO |
| S24-B | 1 | 7 | 28nm | 26nm | |

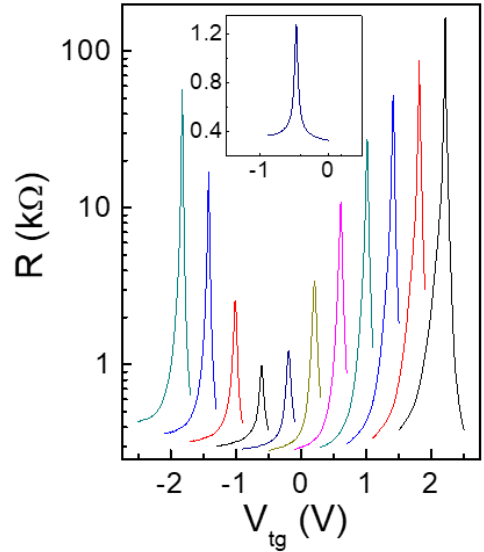

FIG. S2. A semi-log plot of two-terminal resistance $R$ vs. $V_{tg}$ at fixed $V_{bg}$ values. From left to right: $V_{bg}$ changes from 2 V to -2 V in steps of 0.4 V. The peaks correspond to the charge neutrality point of the dual-gated area at different $D$ field. $V_{Si}$ = 40 V. This corresponds to a doping level of ~ 2.2 × $10^{12}$/cm$^2$ in the access region. From the low resistance tail of the traces, we see that the total resistance of the side contacts + the access region $R_c$ is less than 300 Ω, which is negligible compared to the ~ $h/4e^2$ resistance probed in this work. In addition, $R_c$ changes by less than 10 Ω in the temperature range of 0 - 50 K (data not shown). Its minute temperature dependence cannot account for the non-monotonic $T$-dependence shown in Fig. 3 of the text. Inset: Inset: $R$ ($V_{tg}$) in linear scale, $V_{bg}$ = 0.4 V. The half width at half maximum yields a density inhomogeneity of $\delta n$ = 6 × $10^9$ cm$^{-2}$. $T$ = 0.3 K. From device N01-A.

## 2. Additional data on the metallic phase

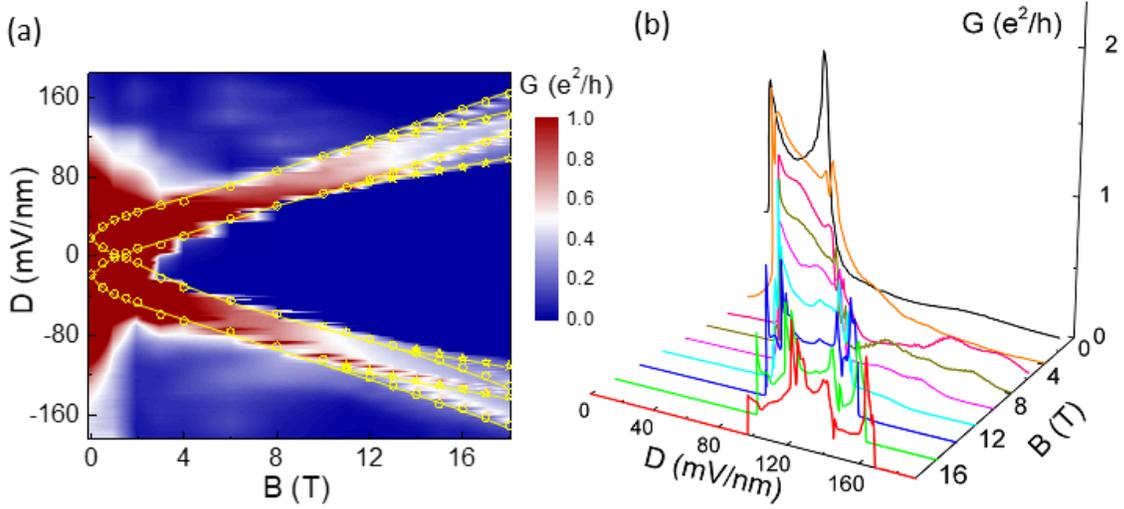

FIG. S3. (a) $G$ ($D$, $B_\perp$) in device N01-A in a false color map constructed using $G$ ($D$) data obtained at fixed $B_\perp$'s similar to the red traces in Fig. 2(a) of the main text. $T$ = 20 mK. The symbols follow that of Fig. 2(a). (b) Stacked $G$ ($D$) traces in the range of 0 to 18 T in steps of 2 T from (a).

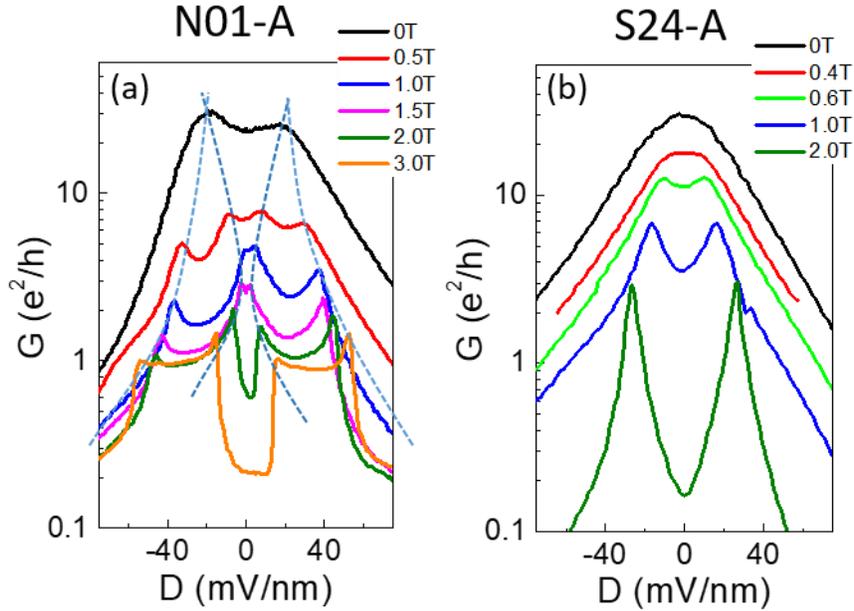

FIG. S4. Comparison of $G$ ($D$) in devices N01-A (a) and S24-A (b) at selected $B_\perp$'s as labeled in the graphs. The blue dashed lines in (a) mark the low $B$-field phase boundaries plotted in Fig. 2(b) of the text. $T$ = 0.3 K and 1.4 K in (a) and (b) respectively.

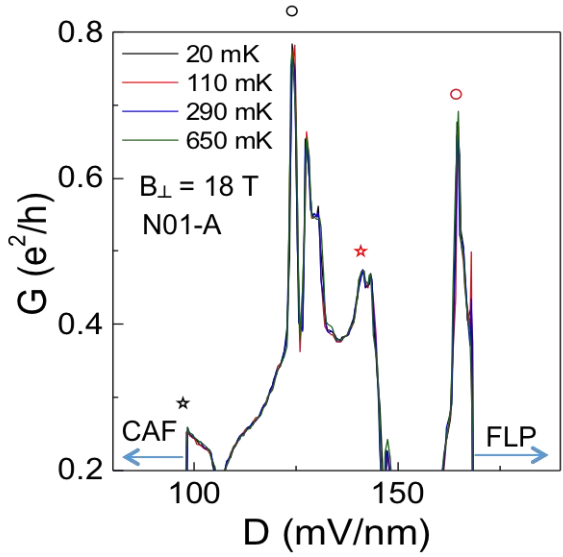 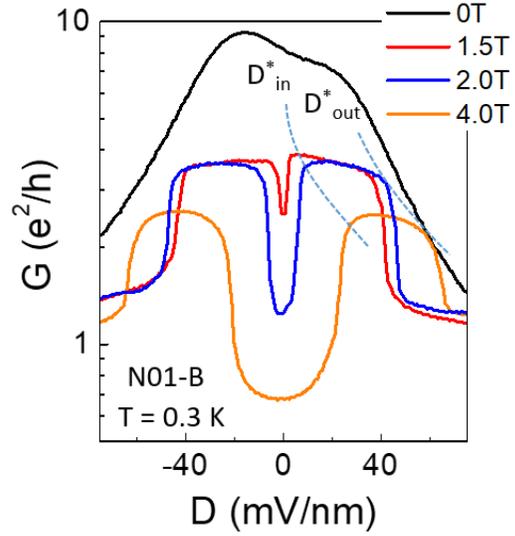

FIG. S5. $G$ ($D$) of $v$ = 0 in device N01-A at $B_\perp$ = 18 T and a few measurement temperatures showing very little temperature dependence of the M phase in this range.

Fig S6. $G$ ($D$) of $v$ = 0 in device N01-B at selected $B_\perp$'s as labeled in the graph. Transitions occur at $D^*_{in/out}$ similar to those in N01-A although the conductance peaks are rounded in this device. N01-B is about 10 μm away from N01-A on the same stack.

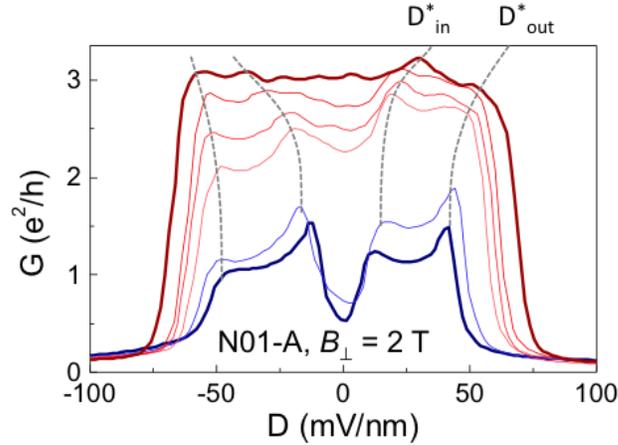

FIG. S7. $G$ ($D$) of $v$ = 0 in device N01-A at $B_\perp$ = 2 T and selected $B_{tot}$'s. From blue to red: $B_{tot}$ = 4, 10, 16, 18, 20, 25 T. The M phase persists to large $B_{tot}$, the conductance of which increases with increasing $B_{tot}$ and is approximately the same as that of the FM phase at large $B_{tot}$. The gray dashed lines mark the boundaries used to construct the phase diagram in Fig 2(c) in the main text.

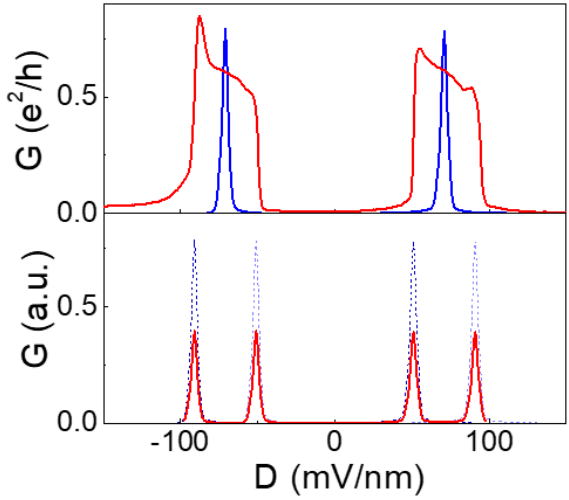

FIG. S8 (a) Upper panel: Conductance vs displacement field $G(D)$ in devices N01-A (red trace, $T = 0.3$ K) and S24-B (blue trace, $T = 1.4$ K). $B_\perp = 8$ T. Lower panel plots the blue trace from upper panel shifted by ±20 mV/nm respectively in dashed blue and their sum in solid red. The summed conductance remains close to zero in the M region since both the CAF and the FLP phases are insulators whereas the measured conductance in N01-A (red trace in the upper panel) is high.

## 3. Additional data on the temperature dependence of the FM and CAF phases

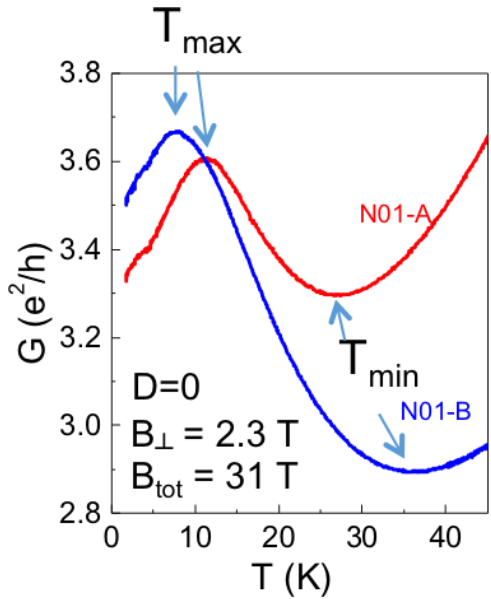

FIG S9. Comparison of $G(T)$ in devices N01-A (red trace) and N01-B (blue trace) in the FM (QSH) phase. $L = 1$ μm, $W = 6$ μm in N01-A and $L = 3$ μm, $W = 6.6$ μm in N01-B. Both traces show qualitatively similar behavior but the temperature at which $G(T)$ reaches maximum or minimum is different for the two devices. This suggests that neither $T_{max}$ nor $T_{min}$ is an intrinsic energy scale. The longer metallic stretch in N01-B is consistent with a weaker bulk contribution in this device because of its dimensions.

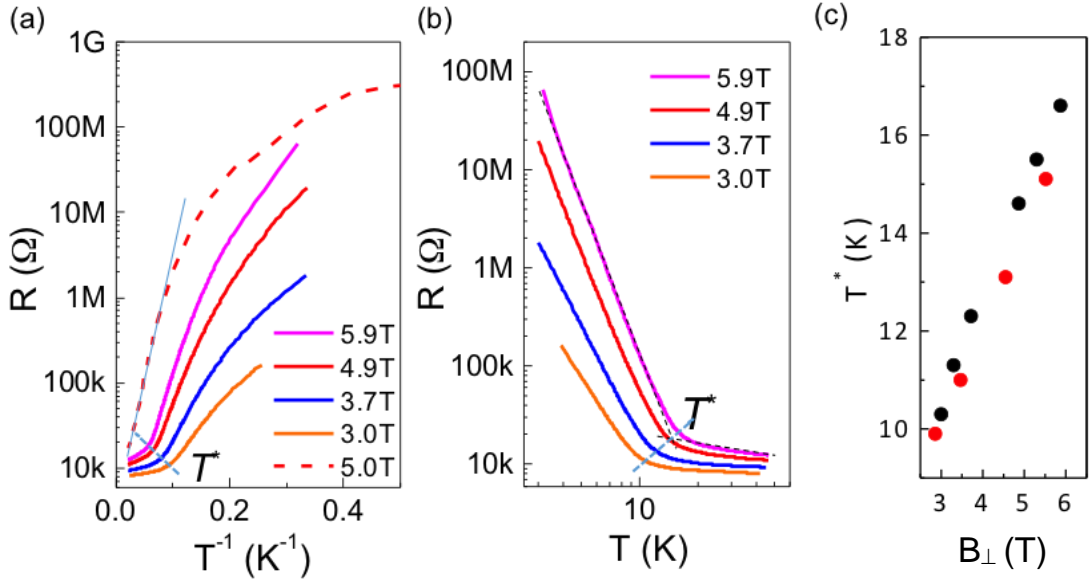

FIG. S10. Temperature dependence of the $v = 0$ CAF state in device S24-A. (a) Arrhenius plot at $B_\perp$'s as labeled in the plot. Solid traces are in the CAF phase at $D = 0$. Below $T = T^*$ defined in (b), $R$ increases rapidly with increasing $T$ but not in a thermally activated manner as the traces curve continuously on the Arrhenius plot. We note that Joule heating or noise is not to blame since this curving behavior is seen at both high and low resistance values. Similar behavior is also observed by Maher et al in Ref. [18] of the text. The dashed trace plots $R(T)$ data in the FLP phase as a comparison. The FLP phase exhibits the expected thermally activated behavior over wide temperature and resistance ranges. A linear fit yields a gap of 150 K at $D = 0.17$ V/nm. (b) The same CAF phase data plotted as a function of $T$ in a log-log scale. Both the high and low-$T$ regimes are well described by power law trend lines that intersect at $T^*$. (c) $T^*$ vs $B_\perp$ in device S24-A (black symbols) and S24-B (red symbols). The linear dependence of $T^*$ on $B_\perp$ points to its correlation with the formation energy of the CAF state. We associate $T^*$ with the onset temperature of the CAF phase. The magnitude of $T^*$ is consistent with the onset temperature of the FM state discussed in Fig. 3 (c) of the text. This apparent power law $T$-dependence of the CAF state requires further studies to clarify. $G(T)$ of data shown here is used to approximate the bulk conduction of the FM (QSH) state possessing the same $D^*$ and is subtracted from measurements in Fig. 3(d) of the text to yield the edge state conduction shown in Fig. 3(e).

17